\documentclass[fleqn,usenatbib]{mnras}

\usepackage{mathptmx}
\usepackage{comment}

\usepackage[T1]{fontenc}
\usepackage{ae,aecompl}

\usepackage{longtable}
\usepackage{placeins}
\usepackage{multicol}
\usepackage{subfigure}

\usepackage{graphicx}	
\usepackage{amsmath}	
\usepackage{amssymb}	
\usepackage{siunitx} 

\title[An accreting SMBH in He 2--10]{Evidence for an accreting massive black hole in He 2-10 from adaptive optics integral field spectroscopy}

\author[R. A. Riffel]{
Rogemar A. Riffel,$^{1,2}$\thanks{E-mail: rogemar@ufsm.br}
\\
$^{1}$Departamento de F\'isica, CCNE, Universidade Federal de Santa Maria, 97105-900, Santa Maria, RS, Brazil\\
$^{2}$Department of Physics \& Astronomy, Johns Hopkins University, Bloomberg Center, 3400 N. Charles St, Baltimore, MD 21218, USA
}

\date{Accepted XXX. Received YYY; in original form ZZZ}

\pubyear{2019}

\begin{document}
\label{firstpage}
\pagerange{\pageref{firstpage}--\pageref{lastpage}}
\maketitle

\begin{abstract}
Henize 2--10 is a blue dwarf galaxy with intense star formation and one the most intriguing question about it is whether or not it hosts an accretingmassive black hole. 
We use H and K-band integral field spectra of the inner 130\,pc$\times$130\,pc  of He 2--10 to investigate the emission and kinematics of the gas at unprecedented spatial resolution.  The observations were done using the Gemini Near-Infrared Integral Field Spectrograph (NIFS) operating with the ALTAIR adaptive optics module and the resulting spatial resolutions are 6.5 pc and  8.6 pc in the K and H bands, respectively. Most of the line emission is due to excitation of the gas by photo-ionization and shocks produced by the star forming regions. In addition, our data provide evidence of emission of gas excited by an active galactic nucleus located at the position of the radio and X-ray sources, as revealed by the analysis of the emission-line ratios. The emission lines from the ionized gas in the field present two kinematic components: one narrow with a velocity field suggesting a disk rotation and a broad component due to winds from the star forming regions. The molecular gas shows only the narrow component. The stellar velocity dispersion map presents an enhancement of about 7 km\,s$^{-1}$ at the position of the black hole, consistent with a mass of $1.5^{+1.3}_{-1.3}\times10^6$\,M$_\odot$. 
\end{abstract}




\section{Introduction}

Supermassive black holes (BH), with masses of $10^6-10^9$\,M$_\odot$, are located at the center of massive galaxies and they are claimed to have a strong impact in the evolution of galaxies \citep[e.g.][]{magorrian98,dimatteo05,kh13,harrison17}.  While they grow, they produce radiative feedback that shapes the star formation of the host galaxy \citep[e.g.][]{jeon12}. Investigate the effect of low-mass ($\sim$10$^{5}-10^{6}$\,M$_{\odot}$) black holes at high-redshift is very difficult from an observational point of view. However, this can be done by observing their analogous in nearby dwarf galaxies \citep[e.g.][]{fs89,reines11,denbrok15}, in particular by using integral field spectroscopy (IFS), which allows the mapping of the gas and stellar distribution and kinematics simultaneously \citep[e.g.][]{vanzi08,cresci10,cresci17,brum19}.

Henize 2--10 (ESO 495- G 021) is one of the most interesting star-forming nearby blue dwarf galaxies, with controversial results  regarding the presence of an accreting black hole \citep{reines11,cresci17}. It has a very high star formation rate and shows signs of a recent interaction \citep{kobulnicky95,reines11}, likely a common event at high z. Indeed, the morphology, metallicity, spectral energy distribution and star--formation rate of a lensed dwarf starburst at $z=1.85$ were shown to resemble local starbursting dwarfs \citep{brammer12}, making this kind of object a particularly promising early-Universe analogue.

\citet{reines11} present optical, near-infrared, radio and X-ray images of He 2--10 obtained with the Hubble Space Telescope (HST), Very Large Array (VLA) and Chandra X-ray Observatory. Based on the ratio between the radio and X-ray luminosities they find that the compact radio source at the dynamical center of the galaxy, and previously reported by \citet{johnson03}, is consistent with an accreting black hole  with a mass of $\sim10^6$\,M$_\odot$.  \citet{reines16} present follow-up X-ray observations of He 2--10 and find that the X-ray source, previously reported, is offset from the radio source and likely an X-ray binary. These deep observations reveal a new X-ray source, co-spatial with the radio source and consistent with a massive BH radiating at 10$^{-6}$  $L_{\rm Edd}$. Very long baseline interferometry observations of He 2-10 at 1.4 GHz reveal a compact non-thermal radio source at the location of the putative accreting BH with a physical scale of $\lesssim$3 pc$\times$1 pc \citep{reines12}. However, 
 recent optical integral field spectroscopy (IFS), obtained with the Multi Unit Spectroscopic Explorer (MUSE) on the Very Large Telescope (VLT) at a seeing of $\sim$0\farcs7 reveal that the emission-line ratios do not show evidence of an active galactic nucleus (AGN) in He 2--10 \citep{cresci17}. These authors, reanalyze the X-ray data and suggest that it is consistent with a supernova remnant origin. The emission-line ratios in the near-infrared are also consistent with gas excitation due to young stars, and evidence of supernova driven winds are seen in locations co-spatial with non-thermal radio sources \citep{cresci10}. \citet{nguyen14} use the Gemini-North's Near-Infrared Integral Field Spectrograph (NIFS) adaptive optics K-band observations of He 2--10 at resolution of 0\farcs15  to measure the stellar kinematics by fitting the K-band CO absorption bandheads and place an upper limit to the mass of the BH of $\sim$10$^7$\,M$_\odot$. Atacama Large Millimeter Array  (ALMA) observations of the cold molecular gas, traced CO(1--0) line, reveal a velocity gradient in molecular gas within the inner 70 pc, consistent with solid body rotation and resulting in a lower limit to the dynamical mass of $2.7\times10^6$\,M$_\odot$, which is consistent with the mass of the BH candidate \citep{imara19}. In addition,  \citet{nguyen14} report that the  Br$\gamma$ emission at the location of the BH candidate presents a luminosity consistent with that expected from the observed X-ray emission. 
 
  Here, we investigate the origin of the near-infrared emission lines from the inner $3^{\prime\prime}\times3^{\prime\prime}$ of He 2--10  using H and K-band integral field unit data obtained with the Gemini NIFS. 
  The NIFS data present an angular resolution about 6 times better than that of the  SINFONI observations previously published by \citet{cresci10}, thus the NIFS observations are more suitable for a detailed analysis of the nature of the gas emission and to look for signatures of a black hole. This paper is organised as follows. Section 2 presents the description of the data and data reduction procedure, Section 3 presents maps for the emission-line fluxes, line-ratios and and kinematics. The results are discussed in Sec. 4 and Sec. 5 presents the conclusions. Through this paper, we adopt the distance to He 2--10 of 9\,Mpc \citep{mendez99} and a corresponding physical scale of 43 pc/arcsec.

 \section{The data and data reduction}
 
We downloaded the NIFS data of He 2--10 from the Gemini Science Archive. The H and K-band observations were done in 2007 April under the program GN-2007A-Q-2 (PI: Usuda) using NIFS together with the ALTtitude conjugate Adaptive optics for the InfraRed (ALTAIR). The stellar kinematics and the Br$\gamma$ emission line flux distribution based on the NIFS K-band data were already published by \citet{nguyen14}. \citet{mcgregor03} present a technical description of the instrument, which has a square field of view of $\approx3\farcs0\times3\farcs0$, divided into 29 slices with an 
angular sampling of 0$\farcs$103$\times$0$\farcs$042. The K-band observations were done using the K$_-$G5605 grating and HK$_-$G0603 filter, while the H band data were obtained using the H$_-$G5605 grating and the JH$_-$G0602 filter. The nominal spectral resolving power for both bands is $R\approx5290$. 

The observations followed the standard Object-Sky-Object dither sequence, with off-source sky positions, and individual exposure times of 300 sec in the K band and 180 sec in the H band. The total on-source exposure times are 45 min and 36 min for the K and H bands, respectively.  The H-band spectra are centred at 1.65 $\mu$m, covering the spectral range from 1.48 to 1.80\,$\mu$m. The K-band data cover the 2.01--2.43\,$\mu$m spectral region and the spectra are centred at 2.20\,$\mu$m.

We followed the standard steps of NIFS data reduction \citep[e.g.,][]{rogemar4051} using the {\sc nifs} package which is part of {\sc gemini iraf} package. The data reduction procedure includes   
the  trimming of the images, flat-fielding, cosmic ray rejection, sky subtraction, wavelength and s-distortion calibrations, removal of the  telluric absorptions, flux calibration by interpolating a blackbody function  to the spectrum of the telluric standard star and construction of data cubes with an angular sampling of 0\farcs05$\times$0\farcs05 for each on-source frame. These individual data cubes were then median combined using a sigma clipping algorithm in order to eliminate bad pixels and taking as reference the position of the  continuum peak to perform the astrometry among the single-exposure data cubes. 

After the data reduction, we partially apply the data treatment procedure described in \citet{menezes14} using {\sc python} and {\sc idl} routines. First, the data cubes are resampled to 0\farcs025 width spaxels using a linear interpolation. Then, we perform a spatial filtering of the data to remove high-frequency noise. We use the {\sc bandpass$_-$filter.pro} IDL routine using a  Butterworth spatial filter and adopting a cutoff frequency of 0.15 Nyquist. 
Finally,  data cubes were resampled back to 0\farcs05 width spaxels.

In order to estimate the angular resolution, we follow \citet{nguyen14} and measure the full-width at half maximum (FWHM) of the flux distribution for the brigthest stellar cluster (identified as CLTC1 by these authors) using H and K continuum images. The continuum images were constructed by computing the average of the fluxes within a spectral window of 100\,\AA, centred at 1.65 and 2.15\,$\mu$m. We measure a FWHM$\approx$0\farcs22 for the K band and 0\farcs25 for the H band. \citet{nguyen14} measured FWHM$\approx$0\farcs19 for CLTC1 using a high resolution image from HST High Resolution Channel (HRC) obtained through the filter F814W.  Thus, the resolution of the NIFS data is  0\farcs15 in the K-band and 0\farcs19 in the H band. These values correspond to linear scales of 6.5 pc  8.6 pc at the distance of the galaxy. The resulting velocity resolution is 45$\pm$5\,km\,s$^{-1}$ for both bands, as measured from the FWHM of typical emission lines of the Ar and ArXe lamps spectra used in the wavelength calibration.

Compared to the seeing limited SINFONI data of He 2--10 used by \citet{cresci10}, the NIFS data present an angular resolution about 6 times better and 75\,\% (30\,\%) higher spectral resolution in the H (K) band than theirs. Similarly, the angular resolution of the NIFS data is almost 7 times smaller than that of the optical MUSE data presented by \citet{cresci17} and the NIFS velocity resolution is as up to 3 times better than that of the blue part of the MUSE spectra. Thus, the NIFS data can provide a mapping of the gas distribution and kinematics of the inner 130\,pc$\times$130\,pc of He 2--10 in unprecedented detail. On the other hand the instruments used in the studies above present a larger field of view than NIFS, providing information about the physical properties of He 2--10 at larger scales.

\begin{figure*}
    \centering
    \includegraphics[width=0.24\textwidth]{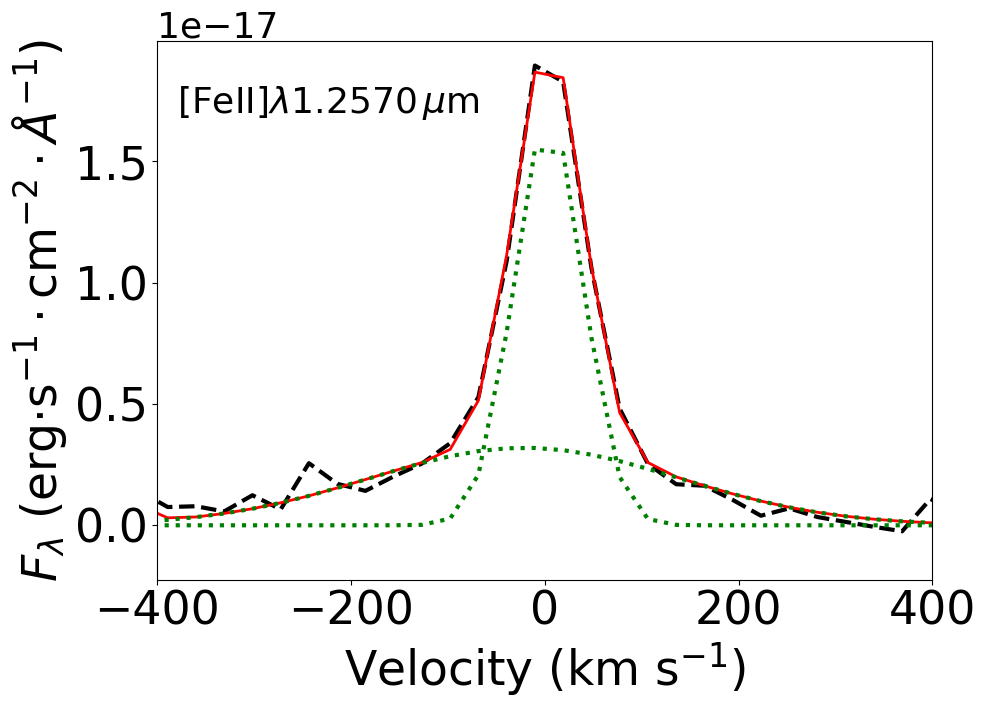}
        \includegraphics[width=0.24\textwidth]{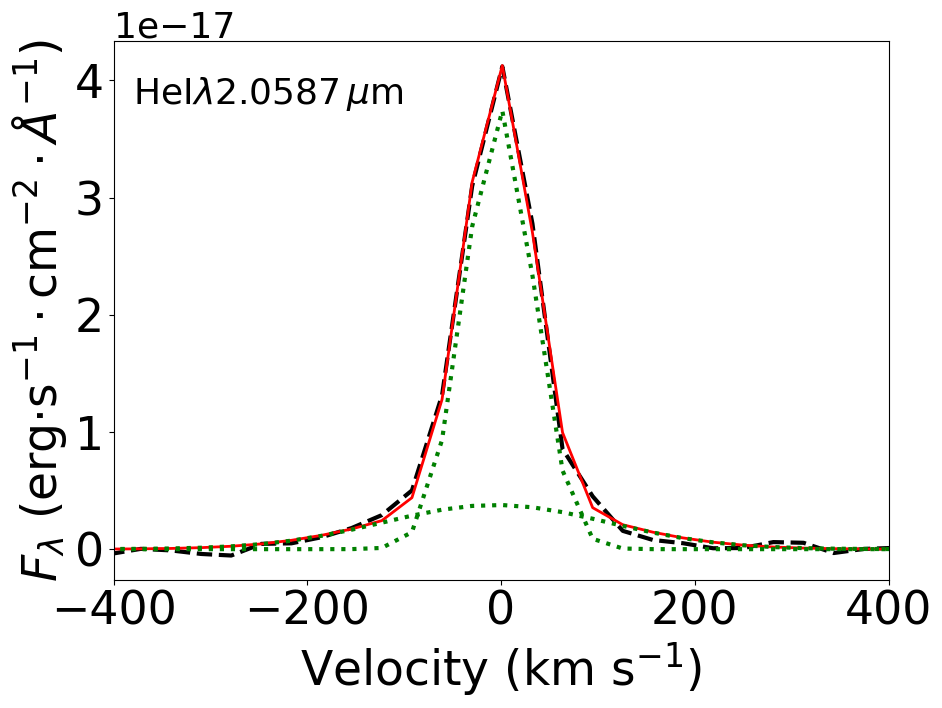}
    \includegraphics[width=0.24\textwidth]{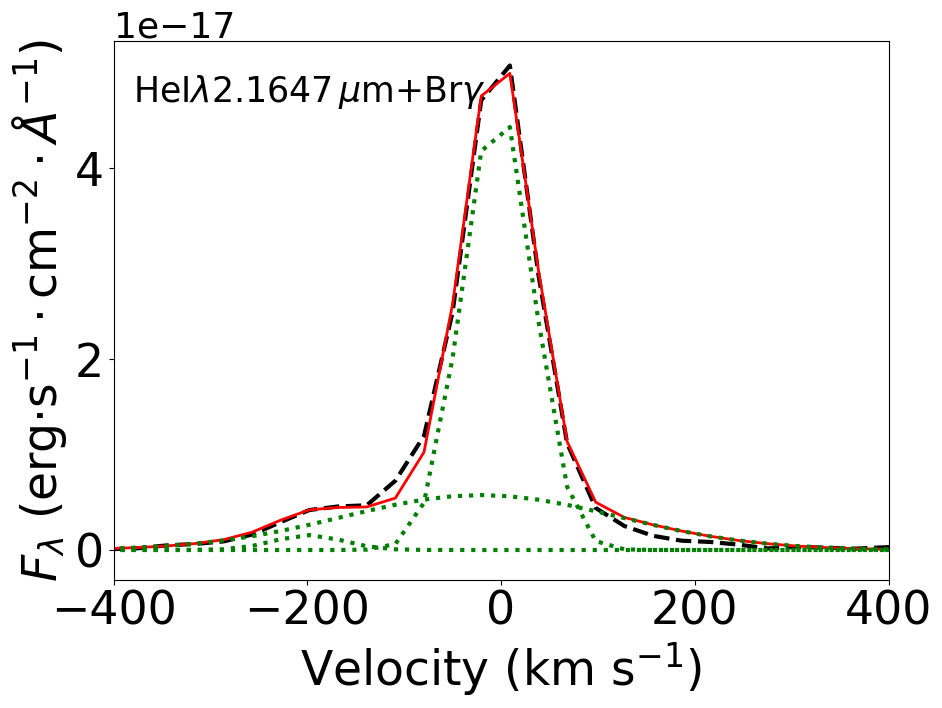}
    \includegraphics[width=0.24\textwidth]{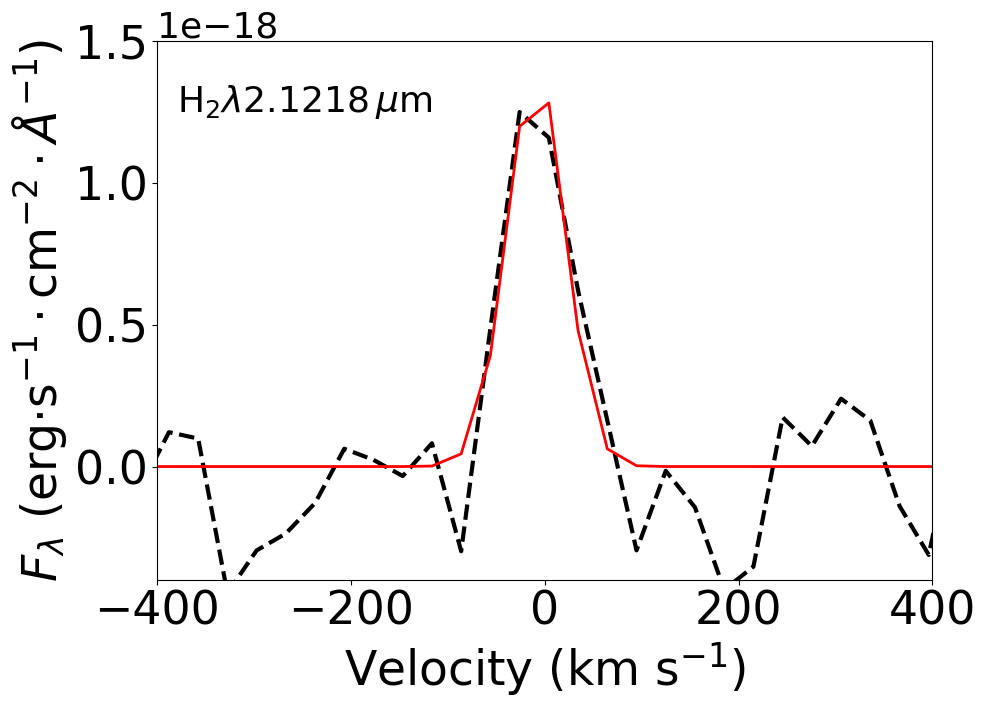}
        \includegraphics[width=0.24\textwidth]{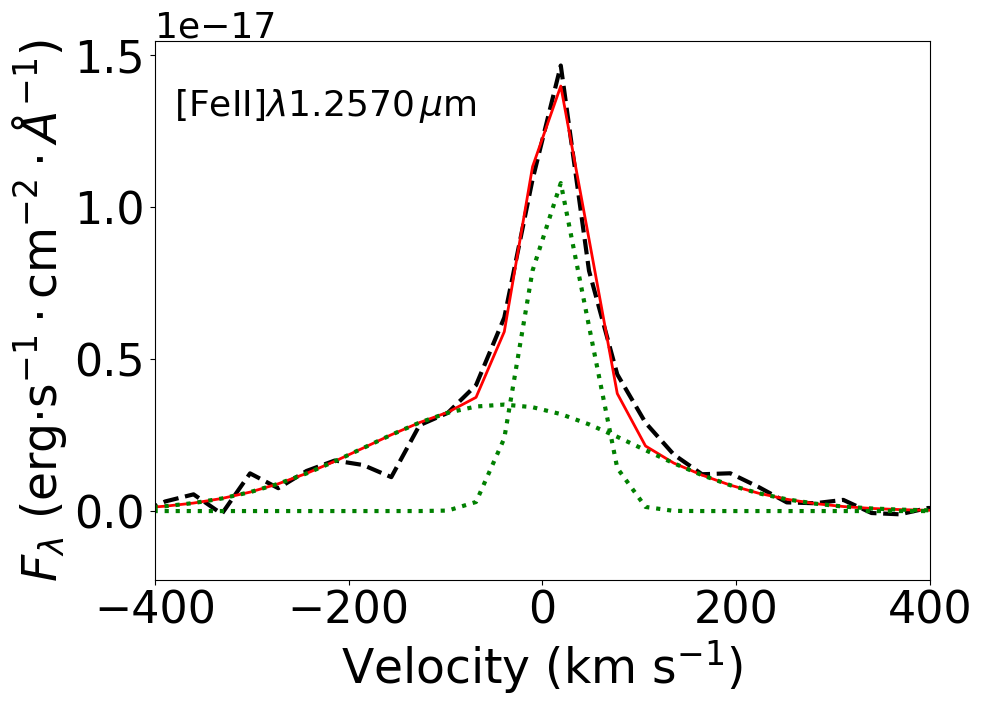}
        \includegraphics[width=0.24\textwidth]{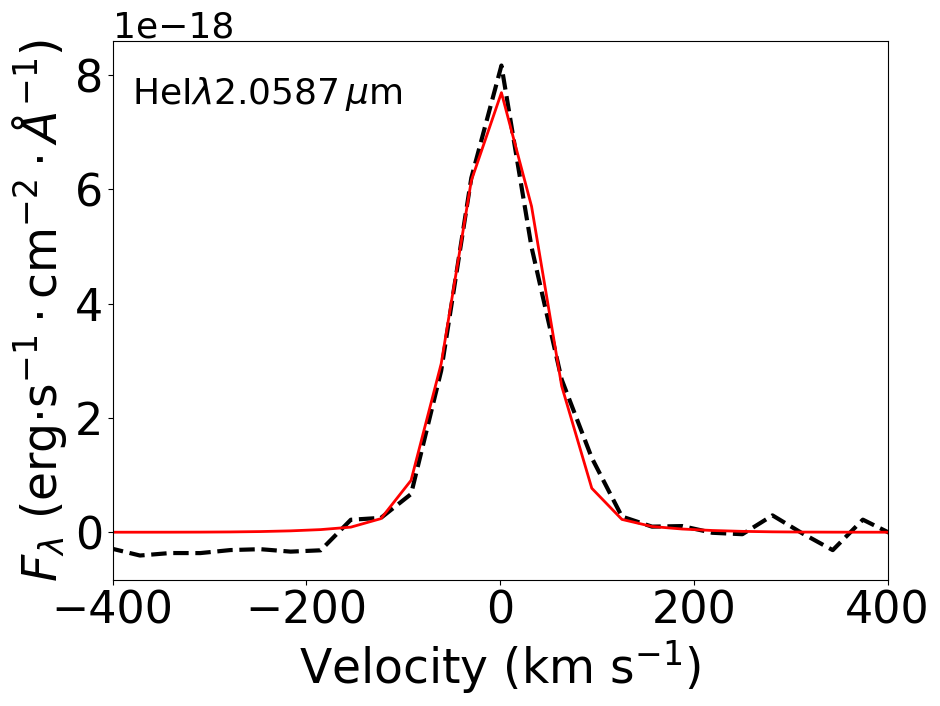}
    \includegraphics[width=0.24\textwidth]{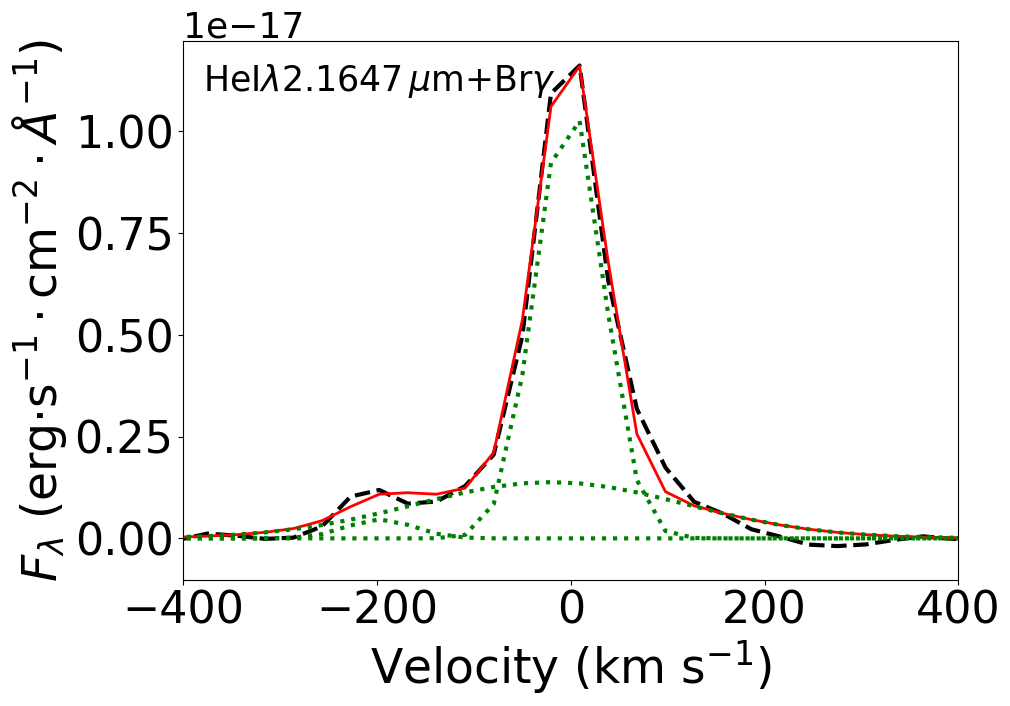}
    \includegraphics[width=0.24\textwidth]{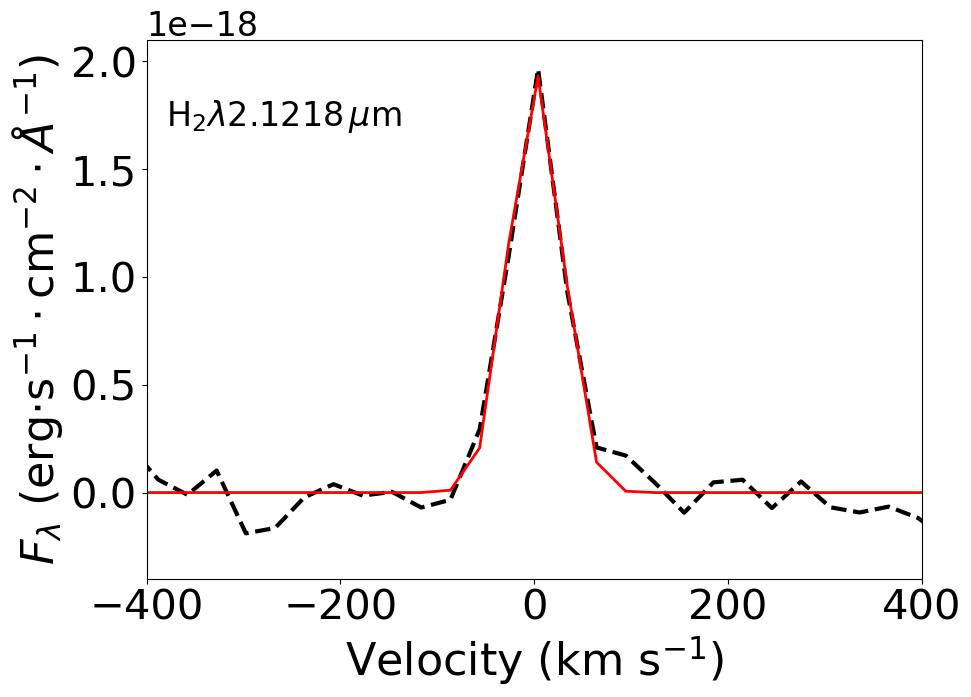}
    \caption{Examples of fits of the emission-line profiles of the [Fe\,{\sc ii}]$\lambda$1.6440\,$\mu$m, He\,{\sc i}$\lambda$2.0587$\,\mu$m, Br$\gamma$ and  H$_2\lambda$2.1218\,$\mu$m at the position of the peak of Br$\gamma$ emission (top panels) and at the location of the BH candidate (bottom panels). The observed profiles are shown as dashed black lines, the model as red continuous lines and the green dotted lines show the individual Gaussian components.}
    \label{fits}
\end{figure*}

  \section{Measurements}

We use the  {\it IFSCube} python package\footnote{https://ifscube.readthedocs.io} to fit the emission line profiles by Gaussian curves. In most locations, the emission lines from the ionized gas present a broad base, besides a narrow component, while the H$_2$ lines present only the narrow component. The [Fe\,{\sc ii}]$\lambda$1.6440\,$\mu$m, He\,{\sc i}$\lambda$2.0587\,$\mu$m and Br$\gamma$ emission lines were fitted using two Gaussian components, while the H$_2\lambda$2.1218\,$\mu$m is well reproduced by a single Gaussian function in all locations. The velocity dispersion of the broad Gaussian components was constrained to be larger than 100 km\,s$^{-1}$, based on visual inspection of the line profiles. If the amplitude of the broad component is smaller than 3 times the standard deviation of the adjacent continuum, the corresponding line profile is fitted by a single Gaussian function. The Br12 emission line, used to estimate the dust extinction, was fitted by a single Gaussian in all locations, as the broad component (if present) is very faint, with amplitude smaller than 3 times the standard deviation of the continuum. In addition, during the fit of the Br$\gamma$ emission line, we include a Gaussian curve to fit the He\,{\sc i}$\lambda$2.1647\,$\mu$m, which is detected and blended with the Br$\gamma$ in some locations. Examples of the observed line profiles and corresponding fits are shown in Fig.~\ref{fits}.

Although, \citet{nguyen14} has already presented stellar kinematics maps for He 2--10 based on the fitting of the K-band NIFS data, our data treatment improves signal-to-noise of the spectra allowing measurements spaxel-by-spaxel, rather than using spatial binned spectra as done by these authors. Thus, we present new measurements for the stellar kinematics based on the treated data cube.  The  stellar line-of-sight velocity distribution is measured by fitting the CO absorption bandheads at $\sim$2.29--2.40\,$\mu$m  using the Pixel-Fitting {\sc ppxf} routine \citep{cappelari04,cappelari17}. 
The spectra of the Gemini library of late spectral type stars observed with the Gemini Near-Infrared Spectrograph (GNIRS) IFU and  NIFS \citep{winge09} are used as spectral templates. This library includes spectra of stars with spectral types from F7 to M5, observed at a very similar spectral resolution than that of the K-band cube of He 2--10.

 \section{Results}
 
\begin{figure*}
    \centering
    \includegraphics[width=1\textwidth]{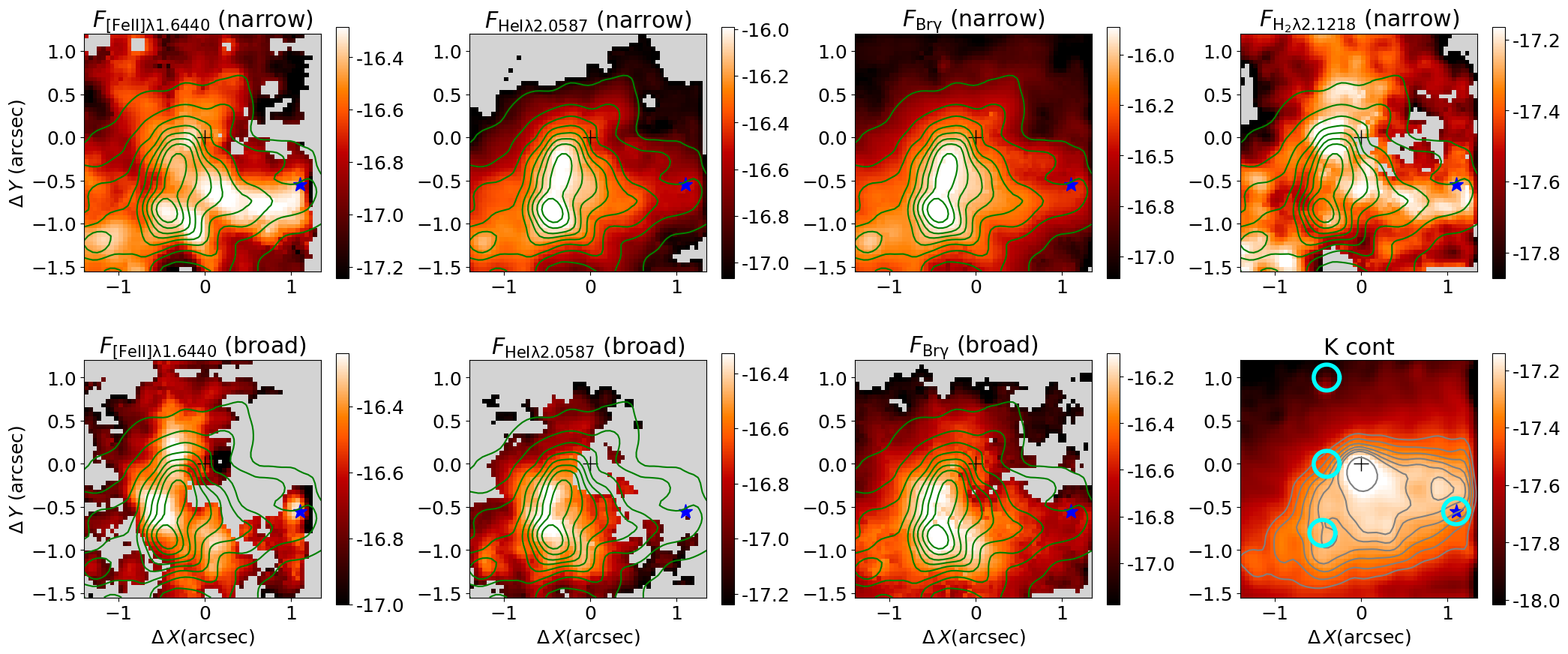}
    \caption{Flux distributions for the narrow and broad line components and K-band continuum image. The green contours overlaid on all maps, but the K-band image, are from the Br$\gamma$ narrow component, while the gray contours are from the K-band image. The cross marks the position of the stellar cluster CLTC1 and the star shows the position of the BH candidate. The color bars show the fluxes in logarithmic scale of erg\,s$^{-1}$\,cm$^{-2}$\,spaxel$^{-1}$ (the size of the spaxels is 0\farcs05$\times$0\farcs5). Gray regions correspond to masked locations where the amplitude of the fitted Gaussian curve is smaller than 3 times the standard deviation of the continuum next to the corresponding emission line.  The cyan circles overploted to the continuum image correspond to the apertures used to extract the spectra discussed in Sec.~\ref{discussion}. In all maps, north is up and east is to the left. }
    \label{flux}
\end{figure*}

Figure~\ref{flux} presents the flux distributions for the [Fe\,{\sc ii}]$\lambda$1.6440\,$\mu$m, He\,{\sc i}$\lambda$2.0587\,$\mu$m, Br$\gamma$  and  H$_2\lambda$2.1218\,$\mu$m emission lines, as well as the K-band continuum.  The positions are given relative to the location of the stellar cluster CLTC1 ($\alpha=$08$^{\rm h}$36$^{\rm m}$15$^{\rm s}$.199, $\delta=-$26$^\circ$24$^\prime$33\farcs62), which is co-spatial with the peak of the near-IR continuum  \citep{reines11,nguyen14}. The star marks the location of BH candidate ($\alpha=$08$^{\rm h}$36$^{\rm m}$15$^{\rm s}$.12, $\delta=-$26$^\circ$24$^\prime$34\farcs1), defined as the position of the X-ray and compact radio souce \citep{reines12,reines16}.  The astrometric uncertainty is $\sim$0\farcs15 (the NIFS angular resolution). The  Br$\gamma$ and He\,{\sc i}$\lambda$2.0587\,$\mu$m show similar flux distributions, with emission observed over most of the NIFS field of view for the narrow component, while the broad component is detected mainly to the southeast. The highest flux levels are mainly seen in knots of emission to the south--southeast, which are surrounded by moderate line emission that extends to locations close to the position of the BH candidate. The faintest emission is observed at the north side of the field of view.  

The H$_2\lambda$2.1218\,$\mu$m and the narrow component of the [Fe\,{\sc ii}]$\lambda$1.6440\,$\mu$m  show overall  similar flux distributions and distinct to that of Br$\gamma$. Strong [Fe\,{\sc ii}] and H$_2$ emission is seen to the southwest and some knots of emission seem to surround the locations of the strongest Br$\gamma$ emission regions. The highest fluxes for the broad [Fe\,{\sc ii}] component is seen between the two knots delineated by the strongest Br$\gamma$ emission. In addition, a secondary peak  is observed at the location of the BH candidate. 

\begin{figure*}
    \centering
    \includegraphics[width=0.82\textwidth]{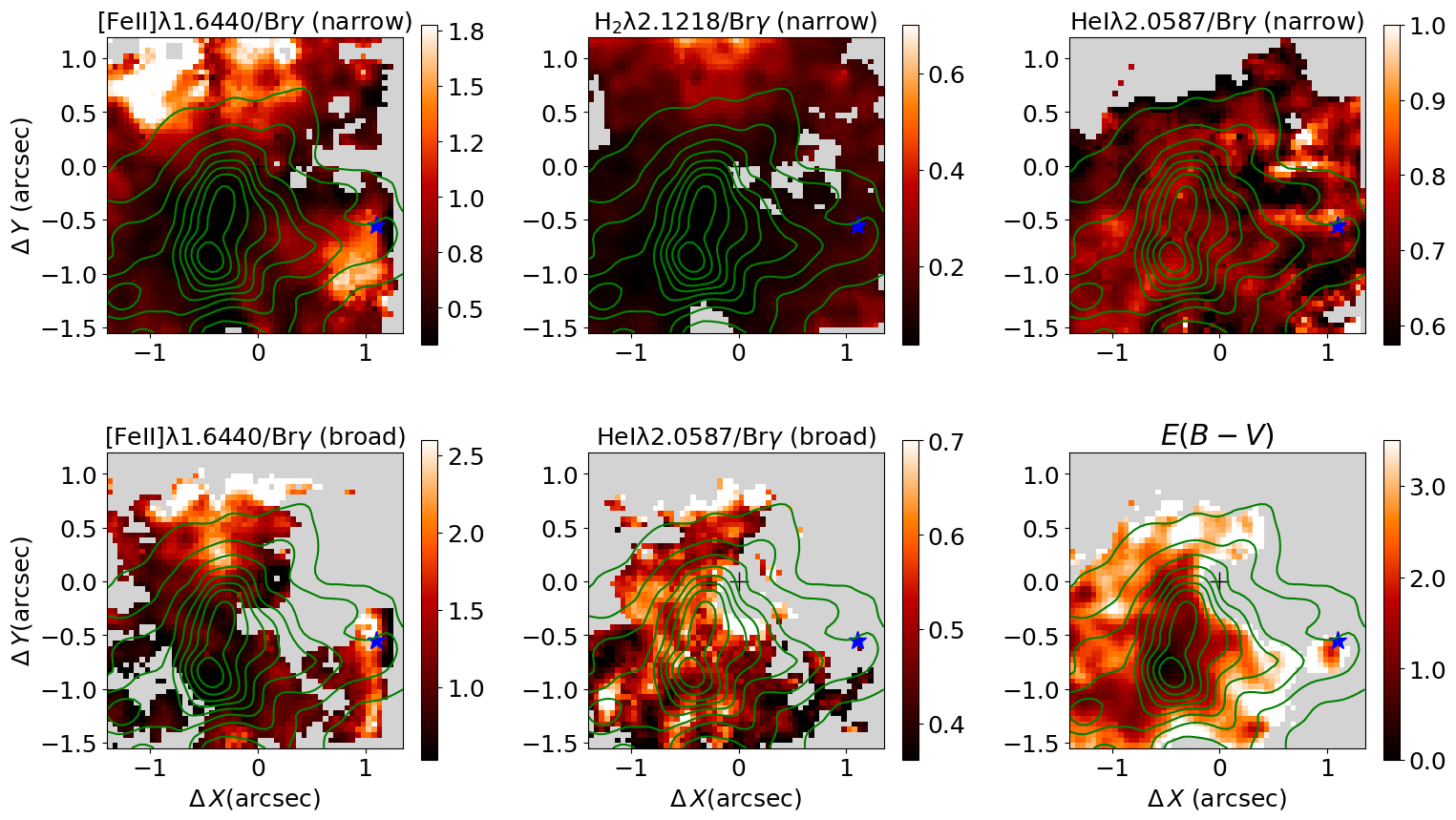}
    \caption{Emission-line ratio maps for He 2--10. Labels are the same as in Fig.\,\ref{flux}.}
    \label{ratio}
\end{figure*}

Emission-line ratio maps are shown in Figure~\ref{ratio}, as well as an $E(B-V)$ map. We estimate the  $E(B-V)$ values by 
\begin{equation}
E(B-V)=10.6\,{\rm log}\left(\frac{F_{\rm Br\gamma}/F_{\rm Br12}}{5.29}\right),
\end{equation}
where $F_{\rm Br\gamma}$ and $F_{\rm Br12}$ are the observed fluxes of Br$\gamma$ and Br12 measured in each spaxel. We use the \citet{ccm89} extinction law, $R_V=3.1$  and adopt the theoretical ratio between ${\rm Br\gamma}$ and ${\rm Br12}$ of 5.29, calculated by \citet{hummer87} for the case B recombination, electron temperature of 10$^4$ K and electron density of 10$^4$ cm$^ {-3}$.  The resulting $E(B-V)$ map is shown in the bottom-right panel of Fig.~\ref{ratio}, with values ranging from $\sim$0.5--3.0 in good agreement with the range of values presented by \citet{cresci10}.

The emission-line ratios shown in Fig.~\ref{ratio} are constructed after correction of the line fluxes using the median value of $\langle E(B-V)\rangle =2.4$ and the extinction law of \citet{ccm89}. The [Fe\,{\sc ii}]/Br$\gamma$ and H$_2$/Br$\gamma$ ratios are commonly used to investigate the origin of the [Fe\,{\sc ii}] and H$_2$ emission \citep{reunanen02,ardila04,ardila05,rogerio13,colina15}. AGN-dominated sources present $-0.3\leq$log\,[Fe\,{\sc ii}]/Br$\gamma\leq1.5$ and  $-0.3\leq$log\,H$_2$/Br$\gamma\leq0.9$, SNe-dominated sources show $0.2\leq$log\,[Fe\,{\sc ii}]/Br$\gamma\leq1.2$ and  $-0.4\leq$log\,H$_2$/Br$\gamma\leq0.4$ and gas excitation by young stars result in $-0.4\leq$log\,[Fe\,{\sc ii}]/Br$\gamma\leq0.4$ and  $-1.2\leq$log\,H$_2$/Br$\gamma\leq-0.1$ \citep{colina15}. For He 2--10, at most locations [Fe\,{\sc ii}]/Br$\gamma\leq1.0$ for both narrow and broad components. The broad component shows overall higher ratios than the narrow component and the highest values are seen to the north and in regions close to the location of the BH candidate. Similarly,  the  H$_2$/Br$\gamma$ shows small values ($<0.2$) at most locations and some higher values in the northern side of the field of view. As the He has a higher ionization potential than H, the He\,{\sc i}/Br$\gamma$ ratio can be used as a treacer of the ionization distribution. From  Fig.~\ref{ratio}, we note that there is a trend of higher values to the west and at locations close to the BH candidate.

\begin{figure*}
    \centering
    \includegraphics[width=1\textwidth]{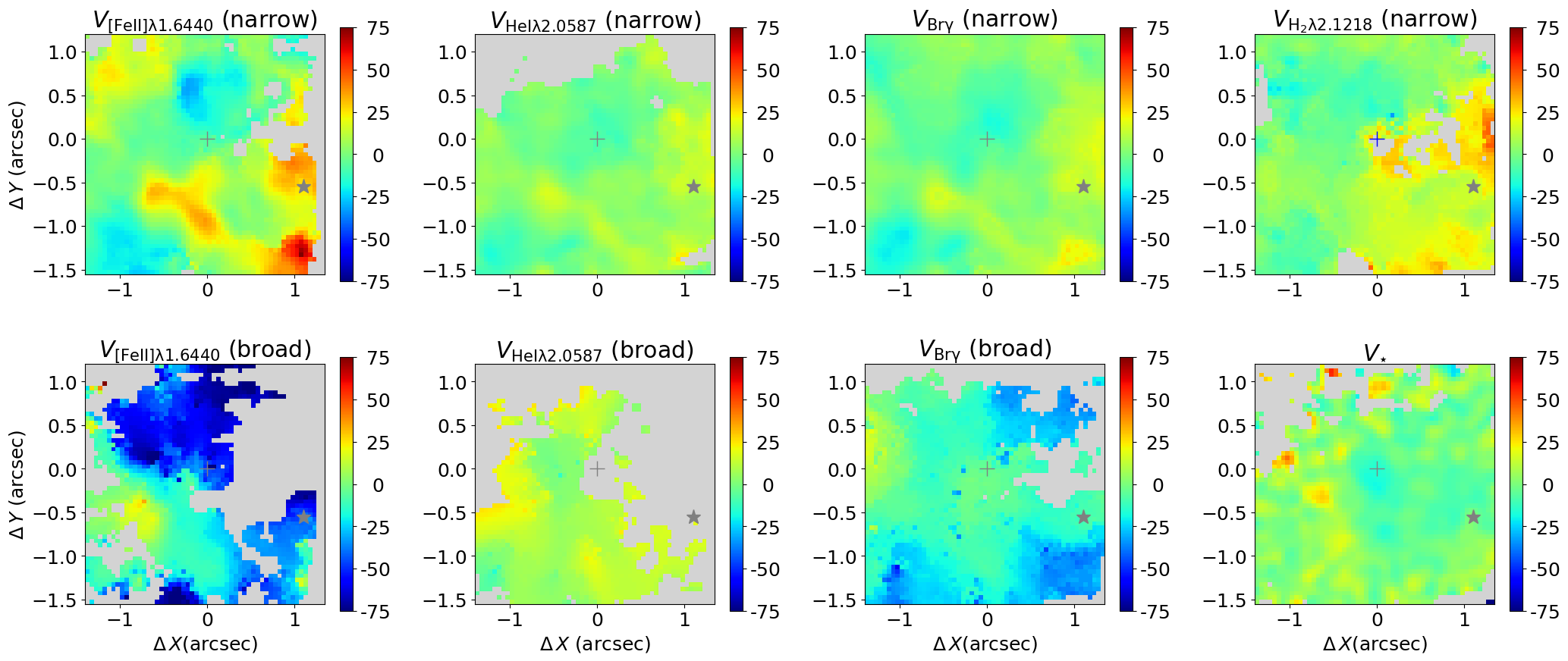}
    \caption{Gas and stellar velocity fields of He 2--10. Labels are the same as in Fig.~\ref{flux}.}
    \label{vel}
\end{figure*}

Figure~\ref{vel} shows the gas velocity maps derived from the centroid of the fitted Gaussian curves, as well as the stellar velocity field derived from the fitting of the CO bands. The observed velocities were corrected to the barycentric  velocity and the systemic velocity of the galaxy -- $V_s=872\pm6$ \citep{marquart07}, is subtracted. The stellar velocity field shows no sign of rotation with velocities smaller than 25 km\,s$^{-1}$ and is in agreement with the map derived by \citet{nguyen14}. On the other hand, the velocity fields for the narrow components of all lines show mostly redshifts to the southwest and blueshifts to the northeast, suggesting a rotation pattern. The maps for the broad components of [Fe\,{\sc ii}]$\lambda$1.6440\,$\mu$m and  Br$\gamma$ show blushifts at most locations, with higher velocities seen for the [Fe\,{\sc ii}]. The  He\,{\sc i}$\lambda$2.0587$\,\mu$m shows velocities close to the systemic velocity of the galaxy.

\begin{figure*}
    \centering
    \includegraphics[width=1\textwidth]{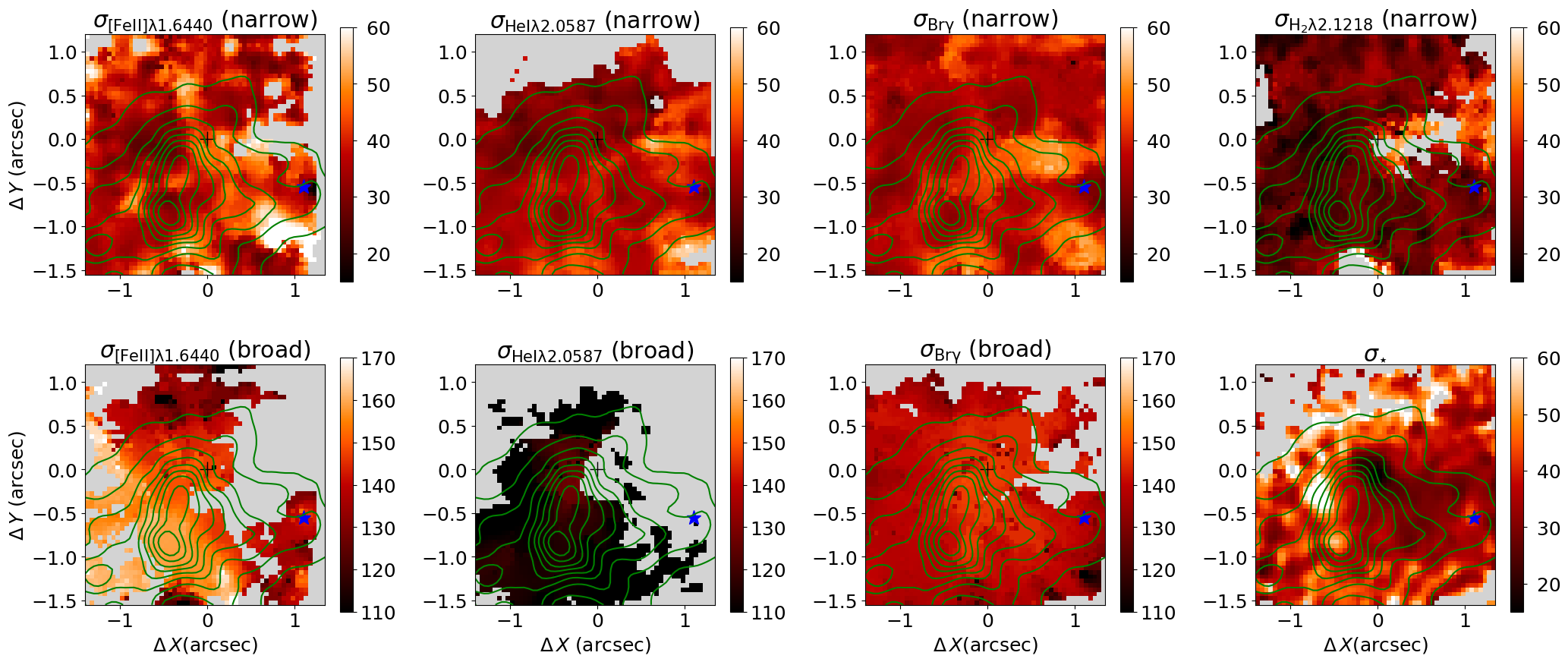}
    \caption{Gas and stellar velocity dispersion maps for He 2--10.  Labels are the same as in Fig.~\ref{flux}.}
    \label{sig}
\end{figure*}

The velocity dispersion ($\sigma$) maps, corrected for the instrumental broadening, are shown in Figure~\ref{sig}. The H$_2$ shows the smallest $\sigma$ values and the highest values for the narrow components (50-60 km\,s$^{-1}$) are seen surrounding the knots of Br$\gamma$ emission. The $\sigma$ maps of the broad components of distinct lines show different range of values. The He\,{\sc i} shows the smallest values ($\sim$120\,km\,s$^{-1}$), followed by Br$\gamma$ ($\sim$140\,km\,s$^{-1}$) and [Fe\,{\sc ii}] presents values of up to 170\,km\,s$^{-1}$.
 
 \section{Discussion}\label{discussion}
 
 The most intriguing question on He 2--10 is whether or not it hosts an accreting massive black hole. \citet{reines11} report that it contains a compact radio source, co-spatial with a hard X-ray source and conclude that the X-ray and radio emission are consistent with an actively accreting black hole with a mass of $\sim10^6$\,M$_\odot$. Follow up radio \citep{reines12} and X-ray \citep{reines16}  observations provide further evidence of an accreting BH in He 2--10, radiating well below its Eddington limit. \citet{nguyen14} derive an upper limit of $\sim$10$^7$\,M$_\odot$ for the BH mass, based on the non-detection of increased stellar velocity dispersion at the location of the BH candidate using Gemini NIFS data. 
 Recent ALMA observations reveal a velocity gradient 
consistent with solid-body rotation and implying in a combined mass
of the star clusters and the BH candidate of $\sim$4$\times10^6$\,M$_\odot$ \citep{imara19}. However, the optical line ratios of He 2--10 do not show sign of gas ionization by an AGN as obtained using seeing limited IFS with the VLT MUSE instrument \citep{cresci17}. 

The high angular and spectral resolution data presented here can be used to investigate whether or not He 2--10 hosts an accreting BH. At the location of the BH candidate, we find significant line emission (see Fig.~\ref{flux}), enhanced emission-line ratios (see Fig.~\ref{ratio}), consistent with those reported for active galaxies \citep[e.g.][]{ardila05,colina15}. In addition, the stellar velocity dispersion map (Fig. \ref{sig}) presents  slightly higher values at the BH position, providing kinematic support for the presence of a compact object. On the other hand, there are no coronal lines or enhanced continuum due to dust emission in the K-band. However, coronal lines and dust emission are not ubiquitous in type 2 AGN \citep{rogerio06,rogerio09_sysp,rogerio11_1157,ardila11,lamperti17}.

The highest [Fe\,{\sc ii}]/Br$\gamma$ values of up to 2.5 are seen close to the location of the BH candidate and to the north of CLTC1, while small values are seen at locations near the knots of Br$\gamma$ emission, consistent with emission of gas excited by young stellar clusters. A similar behaviour is seen for the H$_2$/Br$\gamma$ ratio.  The  high [Fe\,{\sc ii}]/Br$\gamma$ ratios are consistent with an AGN origin, but could also be originated by shocks due to SN winds \citep{colina15}. \citet{cresci10} do also report enhanced line ratios and suggest that they are due to shocks produced by SN winds.  Indeed, the detection of broad line components support the presence of winds associated to the star-forming regions. We find a gradient in the centroid velocity and $\sigma$ among distinct ionization degrees. The He\,{\sc i} shows velocities close to the systemic velocity of the galaxy and the smallest $\sigma$ values, blueshifts are seen in Br$\gamma$ and moderate $\sigma$ values and the highest blueshifts and $\sigma$ values are seen in [Fe\,{\sc ii}]. The He\,{\sc i}$\lambda$2.0587$\,\mu$m line traces the emission of the ionized gas with the highest ionization potential among the detected lines, followed by Br$\gamma$ and [Fe\,{\sc ii}]$\lambda$1.6440\,$\mu$m. The velocity and $\sigma$ gradients are consistent with a scenario in which the He\,{\sc i} traces the gas emission closer to the young stellar clusters, where the dust has already been swept away by the stellar winds, while the [Fe\,{\sc ii}] traces the `dustier' gas farther away from them, where the winds interact with the interstellar medium. Thus, the [Fe\,{\sc ii}] broad component traces the most turbulent gas in dustier regions, consistent with the highest $\sigma$ values and blueshifted velocities, while the redshifeted counterpart of the wind is partially extincted by dust. This scenario is also consistent with the non-detection of a broad component in the H$_2$ lines, which trace a denser gas phase and the winds may not be powerful enough to significantly affect the kinematics of the dense gas.

As discussed above, the [Fe\,{\sc ii}]/Br$\gamma$ and H$_2$/Br$\gamma$ line ratios alone cannot be used to distinguish between AGN excitation and SN winds as the origin of the line emission. The excitation mechanism of the H$_2$ lines can be further investigated using the  H$_2$ 2--1 S(1)/1--0 S(1) and H$_2$ 1--0 S(2)/1--0 S(0) line ratios \citep[e.g.][]{mouri94,dors12,rogerio13}. For He 2--10, the lines 1--0 S(0), 1--0 S(2) and 2--1 S(1) are weak and we are not able to measure their fluxes spaxel-by-spaxel. Thus, we extract spectra from four positions with respect to CLTC1: (A) the peak of Br$\gamma$ emission, at 0\farcs8 southeast; (B) the peak of the H$_2$ emission, at 0\farcs4 east; (C) region with high H$_2$/Br$\gamma$ and [Fe\,{\sc ii}]/Br$\gamma$ at 1\farcs1 north-northeast, and (D)  position of the BH candidate (1\farcs3 southwest). The spectra were extracted within circular apertures of 0\farcs15 radius, as identified by the cyan circles in Fig.~\ref{flux}, and the corresponding flux ratios are shown in Table~\ref{tabela}.

\begin{figure}
    \centering
    \includegraphics[width=0.4\textwidth]{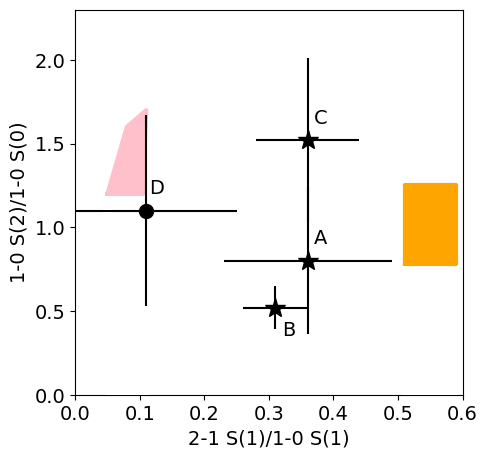}
    \caption{H$_2$ 2--1 S(1)/1--0 S(1) vs. 1--0 S(2)/1--0 S(0) diagram. The points show the observed H$_2$ emission-line ratios for the positions identified in Fig.~\ref{flux} (from A to D). The position D corresponds to the location of the BH candidate. The orange rectangle shows the range of values  predicted by the non-thermal UV excitation models of \citet{black87} and the pink polygon indicates the range of values of the AGN photoionization models from \citet{dors12} for an ionization parameter  $\log U \leq-1.5$.  } 
    \label{h2ratios}
\end{figure}

Figure~\ref{h2ratios} presents the H$_2$ 2--1 S(1)/1--0 S(1) vs. 1--0 S(2)/1--0 S(0) diagram, which can be used to investigate the origin of the H$_2$ emission.   
The line ratios at the positions (A), (B) and (C) are seen between the region occupied by the non-thermal UV excitation models of \citet{black87} -- where the H$_2$ molecule is excited by ultraviolet absorption, followed by fluorescence -- and the region occupied by thermal models \citep[e.g][]{rogerio13} -- the bottom left corner of the S(1)/1--0 S(1) vs. 1--0 S(2)/1--0 S(0) diagram. In addition, the observed line ratios from these three positions are close to the observed data points for star-forming galaxies \citep[e.g.][]{rogerio13}. This suggests that the H$_2$ emission in these regions is produced mainly by fluorescent excitation through absorption of soft-UV photons from the star-forming regions,  but with an additional contribution due to thermal processes (likely shocks due to SN winds), especially for the region (C), that shows the highest 1--0 S(2)/1--0 S(0) ratio and high [Fe\,{\sc ii}]/Br$\gamma$ and H$_2$/Br$\gamma$ values. Indeed, shock models predict 0.3$<$2--1 S(1)/1--0 S(1) $<$ 0.5 \citep{hollenbach89}, consistent with the observed ratios in these locations.  
 On the other hand, at the position of the BH candidate (D), the observed H$_2$ line ratios are very close to the values expected for thermal excitation of the H$_2$. In particular, it is worth noting that the AGN photo-ionization models of \citet{dors12} predict values very close to the observed ratio at position D. Thus, we conclude that the data analysed here provide strong evidence to an accreting black hole in He 2--10, contributing to the gas excitation in its vicinity. The enhancement of the [Fe\,{\sc ii}]/Br$\gamma$  and He\,{\sc i}/Br$\gamma$ line ratios at the position of the BH further supports this conclusion.

 
 \begin{table*}
	\centering
	\caption{H$_2$ 2--1 S(1)/1--0 S(1) and 1--0 S(2)/1--0 S(0) emission line ratios measured the four positions identified as cyan circles in Fig.~\ref{flux}.}
	\label{tabela}
	\begin{tabular}{ccccc} 
		\hline
		Position/Ratio & A (Peak Br$\gamma$) & B (Peak $H_2$)&  C (High H$_2$/Br$\gamma$)&  D (BH candidate) \\
		\hline
		2--1 S(1)/1--0 S(1) &$0.36\pm0.13$&$0.31\pm0.05$&$0.36\pm0.08$&$0.11\pm0.14$ \\
		1--0 S(2)/1--0 S(0) & $0.80\pm0.44$&$0.52\pm0.13$&$1.52\pm0.49$&$1.10\pm0.57$ \\
		\hline
	\end{tabular}
\end{table*}

The stellar velocity dispersion map presents a small enhancement at the location of the BH, which can be used to estimate its mass. The median value of $\sigma_\star$ calculated within a circle with radius $R=$0\farcs15, centred at the position of BH is $<\sigma_\star>_{\rm BH}=39.6\pm3.2$\,km\,s$^{-1}$, while for a ring with inner radius of 0\farcs20 and outer radius of 0\farcs30, the median value is  $<\sigma_\star>_{\rm SP}=32.5\pm2.9$\,km\,s$^{-1}$. If the slightly enhancement in $\sigma_\star$ at the position of the BH  is due its gravitational potential, a rough estimate of BH mass is $1.5^{+1.3}_{-1.3}\times10^6$\,M$_\odot$  as obtained from the Virial theorem by by $M_{\rm BH}=\frac{2\,R(<\sigma_\star>_{\rm BH}^2-<\sigma_\star>_{\rm SP}^2)}{G}$, where $G$ is the Newton's gravitational constant. This value is in agreement with the expected mass ($\sim3\times$10$^6$ M$\odot$) obtained from the relation between the stellar mass and $M_{\rm BH}$ \citep{reines15} by adopting a stellar mass of $\sim$10$^{10}$ M$_\odot$ \citep{nguyen14} and is between the limits calculated by \citet{nguyen14} and \citet{imara19} for He 2--10, as well as between the range of values estimated for BH of NGC\,4395, a well known bulgeless dwarf galaxy hosting a Seyfert 1 nucleus \citep{peterson05,denbrok15,brum19}.

 \section{Conclusion}
 
We use unprecedented high spatial and angular resolution, near-infrared integral field spectra of He 2--10 to investigate the origin of the gas emission and kinematics in the  inner 130\,pc$\times$130\,pc. These data are consistent with the presence of an accreting massiv eblack hole co-spatial with the location of the compact radio and X-ray source previously reported \citep[e.g.][]{reines16}. The main conclusions of this work are:

\begin{itemize}
    \item Most of the near-infrared line emission is due to gas excited by young stars and shocks, likely due to SN winds. But, at the location of the BH, a contribution of an AGN to the gas excitation is strongly supported by the observed line ratios. The [Fe\,{\sc ii}]$\lambda$1.6440\,$\mu$m/Br$\gamma$ and  He\,{\sc i}$\lambda$2.0587$\,\mu$m/Br$\gamma$ present enhanced values at the location of the BH and the H$_2$ 2--1 S(1)/1--0 S(1) and 1--0 S(2)/1--0 S(0) are consistent with values predicted by AGN photo-ionization models. 
    
    \item The emission lines from the ionized gas present two kinematic components: (i) a narrow component with $\sigma<60$\,km\,s$^{-1}$ and velocity field presenting a rotation pattern with redshifts seen to the southwest of the brighetst stellar cluster and blueshifts to the northeast of it and a velocity amplitude of 25 km\,s$^{-1}$. (ii) A broad component seen mostly in blueshifts and with $\sigma>100$\,km\,s$^{-1}$, interpreted as being due to winds from the star forming regions. The molecular gas presents only the disk component. 
    
    \item The stellar velocity field does not present clear signatures of ordered rotation and the stellar velocity dispersion map shows values similar to that of the narrow emission-line component.
    
    \item An enhancement of the stellar velocity dispersion from $32.5\pm2.9$\,km\,s$^{-1}$ to $39.6\pm3.2$\,km\,s$^{-1}$ is observed at the location of the BH, interpreted as being due to motion of stars subjected to the  gravitational potential of a black hole with mass of $1.5^{+1.3}_{-1.3}\times10^6$\,M$_\odot$.
    
\end{itemize}
 
\section*{Acknowledgements}
We thank an anonymous referee for valuable comments that led to  improvements in this paper. 
This study was financed in part by Conselho Nacional de Desenvolvimento Cient\'ifico e Tecnol\'ogico (202582/2018-3, 304927/2017-1 and 400352/2016-8) and Funda\c c\~ao de Amparo \`a pesquisa do Estado do Rio Grande do Sul (17/2551-0001144-9 and 16/2551-0000251-7).

Based on observations obtained at the Gemini Observatory (acquired through the Gemini Observatory Archive and processed using the Gemini IRAF package), which is operated by the Association of Universities for Research in Astronomy, Inc., under a cooperative agreement with the NSF on behalf of the Gemini partnership: the National Science Foundation (United States), National Research Council (Canada), CONICYT (Chile), Ministerio de Ciencia, Tecnolog\'{i}a e Innovaci\'{o}n Productiva (Argentina), Minist\'{e}rio da Ci\^{e}ncia, Tecnologia e Inova\c{c}\~{a}o (Brazil), and Korea Astronomy and Space Science Institute (Republic of Korea). 
This research has made use of NASA's Astrophysics Data System Bibliographic Services. This research has made use of the NASA/IPAC Extragalactic Database (NED), which is operated by the Jet Propulsion Laboratory, California Institute of Technology, under contract with the National Aeronautics and Space Administration.




\bibliographystyle{mnras}
\bibliography{He210} 

\begin{thebibliography}{}
\makeatletter
\relax
\def\mn@urlcharsother{\let\do\@makeother \do\$\do\&\do\#\do\^\do\_\do\%\do\~}
\def\mn@doi{\begingroup\mn@urlcharsother \@ifnextchar [ {\mn@doi@}
  {\mn@doi@[]}}
\def\mn@doi@[#1]#2{\def\@tempa{#1}\ifx\@tempa\@empty \href
  {http://dx.doi.org/#2} {doi:#2}\else \href {http://dx.doi.org/#2} {#1}\fi
  \endgroup}
\def\mn@eprint#1#2{\mn@eprint@#1:#2::\@nil}
\def\mn@eprint@arXiv#1{\href {http://arxiv.org/abs/#1} {{\tt arXiv:#1}}}
\def\mn@eprint@dblp#1{\href {http://dblp.uni-trier.de/rec/bibtex/#1.xml}
  {dblp:#1}}
\def\mn@eprint@#1:#2:#3:#4\@nil{\def\@tempa {#1}\def\@tempb {#2}\def\@tempc
  {#3}\ifx \@tempc \@empty \let \@tempc \@tempb \let \@tempb \@tempa \fi \ifx
  \@tempb \@empty \def\@tempb {arXiv}\fi \@ifundefined
  {mn@eprint@\@tempb}{\@tempb:\@tempc}{\expandafter \expandafter \csname
  mn@eprint@\@tempb\endcsname \expandafter{\@tempc}}}

\bibitem[\protect\citeauthoryear{{Black} \& {van Dishoeck}}{{Black} \& {van
  Dishoeck}}{1987}]{black87}
{Black} J.~H.,  {van Dishoeck} E.~F.,  1987, \mn@doi [\apj] {10.1086/165740},
  \href {https://ui.adsabs.harvard.edu/abs/1987ApJ...322..412B} {322, 412}

\bibitem[\protect\citeauthoryear{{Brammer} et~al.,}{{Brammer}
  et~al.}{2012}]{brammer12}
{Brammer} G.~B.,  et~al., 2012, \mn@doi [\apjl] {10.1088/2041-8205/758/1/L17},
  \href {https://ui.adsabs.harvard.edu/abs/2012ApJ...758L..17B} {758, L17}

\bibitem[\protect\citeauthoryear{{Brum} et~al.,}{{Brum} et~al.}{2019}]{brum19}
{Brum} C.,  et~al., 2019, \mn@doi [\mnras] {10.1093/mnras/stz893}, \href
  {https://ui.adsabs.harvard.edu/abs/2019MNRAS.486..691B} {486, 691}

\bibitem[\protect\citeauthoryear{{Cappellari}}{{Cappellari}}{2017}]{cappelari17}
{Cappellari} M.,  2017, \mn@doi [\mnras] {10.1093/mnras/stw3020}, \href
  {https://ui.adsabs.harvard.edu/abs/2017MNRAS.466..798C} {466, 798}

\bibitem[\protect\citeauthoryear{{Cappellari} \& {Emsellem}}{{Cappellari} \&
  {Emsellem}}{2004}]{cappelari04}
{Cappellari} M.,  {Emsellem} E.,  2004, \mn@doi [\pasp] {10.1086/381875}, \href
  {https://ui.adsabs.harvard.edu/abs/2004PASP..116..138C} {116, 138}

\bibitem[\protect\citeauthoryear{{Cardelli}, {Clayton}  \& {Mathis}}{{Cardelli}
  et~al.}{1989}]{ccm89}
{Cardelli} J.~A.,  {Clayton} G.~C.,   {Mathis} J.~S.,  1989, \mn@doi [\apj]
  {10.1086/167900}, \href
  {https://ui.adsabs.harvard.edu/abs/1989ApJ...345..245C} {345, 245}

\bibitem[\protect\citeauthoryear{{Colina} et~al.,}{{Colina}
  et~al.}{2015}]{colina15}
{Colina} L.,  et~al., 2015, \mn@doi [\aap] {10.1051/0004-6361/201425567}, \href
  {https://ui.adsabs.harvard.edu/abs/2015A&A...578A..48C} {578, A48}

\bibitem[\protect\citeauthoryear{{Cresci}, {Vanzi}, {Sauvage}, {Santangelo}  \&
  {van der Werf}}{{Cresci} et~al.}{2010}]{cresci10}
{Cresci} G.,  {Vanzi} L.,  {Sauvage} M.,  {Santangelo} G.,   {van der Werf} P.,
   2010, \mn@doi [\aap] {10.1051/0004-6361/201014864}, \href
  {https://ui.adsabs.harvard.edu/abs/2010A&A...520A..82C} {520, A82}

\bibitem[\protect\citeauthoryear{{Cresci}, {Vanzi}, {Telles}, {Lanzuisi},
  {Brusa}, {Mingozzi}, {Sauvage}  \& {Johnson}}{{Cresci}
  et~al.}{2017}]{cresci17}
{Cresci} G.,  {Vanzi} L.,  {Telles} E.,  {Lanzuisi} G.,  {Brusa} M.,
  {Mingozzi} M.,  {Sauvage} M.,   {Johnson} K.,  2017, \mn@doi [\aap]
  {10.1051/0004-6361/201730876}, \href
  {https://ui.adsabs.harvard.edu/abs/2017A&A...604A.101C} {604, A101}

\bibitem[\protect\citeauthoryear{{Di Matteo}, {Springel}  \& {Hernquist}}{{Di
  Matteo} et~al.}{2005}]{dimatteo05}
{Di Matteo} T.,  {Springel} V.,   {Hernquist} L.,  2005, \mn@doi [\nat]
  {10.1038/nature03335}, \href
  {https://ui.adsabs.harvard.edu/abs/2005Natur.433..604D} {433, 604}

\bibitem[\protect\citeauthoryear{{Dors}, {Riffel}, {Cardaci}, {H{\"a}gele},
  {Krabbe}, {P{\'e}rez-Montero}  \& {Rodrigues}}{{Dors} et~al.}{2012}]{dors12}
{Dors} Oli~L. J.,  {Riffel} R.~A.,  {Cardaci} M.~V.,  {H{\"a}gele} G.~F.,
  {Krabbe} {\'A}.~C.,  {P{\'e}rez-Montero} E.,   {Rodrigues} I.,  2012, \mn@doi
  [\mnras] {10.1111/j.1365-2966.2012.20600.x}, \href
  {https://ui.adsabs.harvard.edu/abs/2012MNRAS.422..252D} {422, 252}

\bibitem[\protect\citeauthoryear{{Filippenko} \& {Sargent}}{{Filippenko} \&
  {Sargent}}{1989}]{fs89}
{Filippenko} A.~V.,  {Sargent} W. L.~W.,  1989, \mn@doi [\apjl]
  {10.1086/185472}, \href
  {https://ui.adsabs.harvard.edu/abs/1989ApJ...342L..11F} {342, L11}

\bibitem[\protect\citeauthoryear{{Harrison}}{{Harrison}}{2017}]{harrison17}
{Harrison} C.~M.,  2017, \mn@doi [Nature Astronomy] {10.1038/s41550-017-0165},
  \href {https://ui.adsabs.harvard.edu/abs/2017NatAs...1E.165H} {1, 0165}

\bibitem[\protect\citeauthoryear{{Hollenbach} \& {McKee}}{{Hollenbach} \&
  {McKee}}{1989}]{hollenbach89}
{Hollenbach} D.,  {McKee} C.~F.,  1989, \mn@doi [\apj] {10.1086/167595}, \href
  {https://ui.adsabs.harvard.edu/abs/1989ApJ...342..306H} {342, 306}

\bibitem[\protect\citeauthoryear{{Hummer} \& {Storey}}{{Hummer} \&
  {Storey}}{1987}]{hummer87}
{Hummer} D.~G.,  {Storey} P.~J.,  1987, \mn@doi [\mnras]
  {10.1093/mnras/224.3.801}, \href
  {https://ui.adsabs.harvard.edu/abs/1987MNRAS.224..801H} {224, 801}

\bibitem[\protect\citeauthoryear{{Imara} \& {Faesi}}{{Imara} \&
  {Faesi}}{2019}]{imara19}
{Imara} N.,  {Faesi} C.~M.,  2019, \mn@doi [\apj] {10.3847/1538-4357/ab16cc},
  \href {https://ui.adsabs.harvard.edu/abs/2019ApJ...876..141I} {876, 141}

\bibitem[\protect\citeauthoryear{{Jeon}, {Pawlik}, {Greif}, {Glover}, {Bromm},
  {Milosavljevi{\'c}}  \& {Klessen}}{{Jeon} et~al.}{2012}]{jeon12}
{Jeon} M.,  {Pawlik} A.~H.,  {Greif} T.~H.,  {Glover} S. C.~O.,  {Bromm} V.,
  {Milosavljevi{\'c}} M.,   {Klessen} R.~S.,  2012, \mn@doi [\apj]
  {10.1088/0004-637X/754/1/34}, \href
  {https://ui.adsabs.harvard.edu/abs/2012ApJ...754...34J} {754, 34}

\bibitem[\protect\citeauthoryear{{Johnson} \& {Kobulnicky}}{{Johnson} \&
  {Kobulnicky}}{2003}]{johnson03}
{Johnson} K.~E.,  {Kobulnicky} H.~A.,  2003, \mn@doi [\apj] {10.1086/378585},
  \href {https://ui.adsabs.harvard.edu/abs/2003ApJ...597..923J} {597, 923}

\bibitem[\protect\citeauthoryear{{Kobulnicky}, {Dickey}, {Sargent}, {Hogg}  \&
  {Conti}}{{Kobulnicky} et~al.}{1995}]{kobulnicky95}
{Kobulnicky} H.~A.,  {Dickey} J.~M.,  {Sargent} A.~I.,  {Hogg} D.~E.,   {Conti}
  P.~S.,  1995, \mn@doi [\aj] {10.1086/117500}, \href
  {https://ui.adsabs.harvard.edu/abs/1995AJ....110..116K} {110, 116}

\bibitem[\protect\citeauthoryear{{Kormendy} \& {Ho}}{{Kormendy} \&
  {Ho}}{2013}]{kh13}
{Kormendy} J.,  {Ho} L.~C.,  2013, \mn@doi [\araa]
  {10.1146/annurev-astro-082708-101811}, \href
  {https://ui.adsabs.harvard.edu/abs/2013ARA&A..51..511K} {51, 511}

\bibitem[\protect\citeauthoryear{{Lamperti} et~al.,}{{Lamperti}
  et~al.}{2017}]{lamperti17}
{Lamperti} I.,  et~al., 2017, \mn@doi [\mnras] {10.1093/mnras/stx055}, \href
  {https://ui.adsabs.harvard.edu/abs/2017MNRAS.467..540L} {467, 540}

\bibitem[\protect\citeauthoryear{{Magorrian} et~al.,}{{Magorrian}
  et~al.}{1998}]{magorrian98}
{Magorrian} J.,  et~al., 1998, \mn@doi [\aj] {10.1086/300353}, \href
  {https://ui.adsabs.harvard.edu/abs/1998AJ....115.2285M} {115, 2285}

\bibitem[\protect\citeauthoryear{{Marquart}, {Fathi}, {{\"O}stlin}, {Bergvall},
  {Cumming}  \& {Amram}}{{Marquart} et~al.}{2007}]{marquart07}
{Marquart} T.,  {Fathi} K.,  {{\"O}stlin} G.,  {Bergvall} N.,  {Cumming} R.~J.,
    {Amram} P.,  2007, \mn@doi [\aap] {10.1051/0004-6361:20078142}, \href
  {https://ui.adsabs.harvard.edu/abs/2007A&A...474L...9M} {474, L9}

\bibitem[\protect\citeauthoryear{{McGregor} et~al.,}{{McGregor}
  et~al.}{2003}]{mcgregor03}
{McGregor} P.~J.,  et~al., 2003, {Gemini near-infrared integral field
  spectrograph (NIFS)}.
Proceedings of the SPIE, pp 1581--1591, \mn@doi{10.1117/12.459448}

\bibitem[\protect\citeauthoryear{{M{\'e}ndez}, {Esteban}, {Filipovi{\'c}},
  {Ehle}, {Haberl}, {Pietsch}  \& {Haynes}}{{M{\'e}ndez}
  et~al.}{1999}]{mendez99}
{M{\'e}ndez} D.~I.,  {Esteban} C.,  {Filipovi{\'c}} M.~D.,  {Ehle} M.,
  {Haberl} F.,  {Pietsch} W.,   {Haynes} R.~F.,  1999, \aap, \href
  {https://ui.adsabs.harvard.edu/abs/1999A&A...349..801M} {349, 801}

\bibitem[\protect\citeauthoryear{{Menezes}, {Steiner}  \& {Ricci}}{{Menezes}
  et~al.}{2014}]{menezes14}
{Menezes} R.~B.,  {Steiner} J.~E.,   {Ricci} T.~V.,  2014, \mn@doi [\mnras]
  {10.1093/mnras/stt2381}, \href
  {https://ui.adsabs.harvard.edu/abs/2014MNRAS.438.2597M} {438, 2597}

\bibitem[\protect\citeauthoryear{{Mouri}}{{Mouri}}{1994}]{mouri94}
{Mouri} H.,  1994, \mn@doi [\apj] {10.1086/174184}, \href
  {https://ui.adsabs.harvard.edu/abs/1994ApJ...427..777M} {427, 777}

\bibitem[\protect\citeauthoryear{{Nguyen}, {Seth}, {Reines}, {den Brok}, {Sand}
   \& {McLeod}}{{Nguyen} et~al.}{2014}]{nguyen14}
{Nguyen} D.~D.,  {Seth} A.~C.,  {Reines} A.~E.,  {den Brok} M.,  {Sand} D.,
  {McLeod} B.,  2014, \mn@doi [\apj] {10.1088/0004-637X/794/1/34}, \href
  {https://ui.adsabs.harvard.edu/abs/2014ApJ...794...34N} {794, 34}

\bibitem[\protect\citeauthoryear{{Peterson} et~al.,}{{Peterson}
  et~al.}{2005}]{peterson05}
{Peterson} B.~M.,  et~al., 2005, \mn@doi [\apj] {10.1086/444494}, \href
  {https://ui.adsabs.harvard.edu/abs/2005ApJ...632..799P} {632, 799}

\bibitem[\protect\citeauthoryear{{Reines} \& {Deller}}{{Reines} \&
  {Deller}}{2012}]{reines12}
{Reines} A.~E.,  {Deller} A.~T.,  2012, \mn@doi [\apjl]
  {10.1088/2041-8205/750/1/L24}, \href
  {https://ui.adsabs.harvard.edu/abs/2012ApJ...750L..24R} {750, L24}

\bibitem[\protect\citeauthoryear{{Reines} \& {Volonteri}}{{Reines} \&
  {Volonteri}}{2015}]{reines15}
{Reines} A.~E.,  {Volonteri} M.,  2015, \mn@doi [\apj]
  {10.1088/0004-637X/813/2/82}, \href
  {https://ui.adsabs.harvard.edu/abs/2015ApJ...813...82R} {813, 82}

\bibitem[\protect\citeauthoryear{{Reines}, {Sivakoff}, {Johnson}  \&
  {Brogan}}{{Reines} et~al.}{2011}]{reines11}
{Reines} A.~E.,  {Sivakoff} G.~R.,  {Johnson} K.~E.,   {Brogan} C.~L.,  2011,
  \mn@doi [\nat] {10.1038/nature09724}, \href
  {https://ui.adsabs.harvard.edu/abs/2011Natur.470...66R} {470, 66}

\bibitem[\protect\citeauthoryear{{Reines}, {Reynolds}, {Miller}, {Sivakoff},
  {Greene}, {Hickox}  \& {Johnson}}{{Reines} et~al.}{2016}]{reines16}
{Reines} A.~E.,  {Reynolds} M.~T.,  {Miller} J.~M.,  {Sivakoff} G.~R.,
  {Greene} J.~E.,  {Hickox} R.~C.,   {Johnson} K.~E.,  2016, \mn@doi [\apjl]
  {10.3847/2041-8205/830/2/L35}, \href
  {https://ui.adsabs.harvard.edu/abs/2016ApJ...830L..35R} {830, L35}

\bibitem[\protect\citeauthoryear{{Reunanen}, {Kotilainen}  \&
  {Prieto}}{{Reunanen} et~al.}{2002}]{reunanen02}
{Reunanen} J.,  {Kotilainen} J.~K.,   {Prieto} M.~A.,  2002, \mn@doi [\mnras]
  {10.1046/j.1365-8711.2002.05181.x}, \href
  {https://ui.adsabs.harvard.edu/abs/2002MNRAS.331..154R} {331, 154}

\bibitem[\protect\citeauthoryear{{Riffel}, {Rodr{\'\i}guez-Ardila}  \&
  {Pastoriza}}{{Riffel} et~al.}{2006}]{rogerio06}
{Riffel} R.,  {Rodr{\'\i}guez-Ardila} A.,   {Pastoriza} M.~G.,  2006, \mn@doi
  [\aap] {10.1051/0004-6361:20065291}, \href
  {https://ui.adsabs.harvard.edu/abs/2006A&A...457...61R} {457, 61}

\bibitem[\protect\citeauthoryear{{Riffel}, {Storchi-Bergmann}, {Winge},
  {McGregor}, {Beck}  \& {Schmitt}}{{Riffel} et~al.}{2008}]{rogemar4051}
{Riffel} R.~A.,  {Storchi-Bergmann} T.,  {Winge} C.,  {McGregor} P.~J.,  {Beck}
  T.,   {Schmitt} H.,  2008, \mn@doi [\mnras]
  {10.1111/j.1365-2966.2008.12936.x}, \href
  {https://ui.adsabs.harvard.edu/abs/2008MNRAS.385.1129R} {385, 1129}

\bibitem[\protect\citeauthoryear{{Riffel}, {Pastoriza}, {Rodr{\'\i}guez-Ardila}
   \& {Bonatto}}{{Riffel} et~al.}{2009}]{rogerio09_sysp}
{Riffel} R.,  {Pastoriza} M.~G.,  {Rodr{\'\i}guez-Ardila} A.,   {Bonatto} C.,
  2009, \mn@doi [\mnras] {10.1111/j.1365-2966.2009.15448.x}, \href
  {https://ui.adsabs.harvard.edu/abs/2009MNRAS.400..273R} {400, 273}

\bibitem[\protect\citeauthoryear{{Riffel}, {Riffel}, {Ferrari}  \&
  {Storchi-Bergmann}}{{Riffel} et~al.}{2011}]{rogerio11_1157}
{Riffel} R.,  {Riffel} R.~A.,  {Ferrari} F.,   {Storchi-Bergmann} T.,  2011,
  \mn@doi [\mnras] {10.1111/j.1365-2966.2011.19061.x}, \href
  {https://ui.adsabs.harvard.edu/abs/2011MNRAS.416..493R} {416, 493}

\bibitem[\protect\citeauthoryear{{Riffel}, {Rodr{\'\i}guez-Ardila}, {Aleman},
  {Brotherton}, {Pastoriza}, {Bonatto}  \& {Dors}}{{Riffel}
  et~al.}{2013}]{rogerio13}
{Riffel} R.,  {Rodr{\'\i}guez-Ardila} A.,  {Aleman} I.,  {Brotherton} M.~S.,
  {Pastoriza} M.~G.,  {Bonatto} C.,   {Dors} O.~L.,  2013, \mn@doi [\mnras]
  {10.1093/mnras/stt026}, \href
  {https://ui.adsabs.harvard.edu/abs/2013MNRAS.430.2002R} {430, 2002}

\bibitem[\protect\citeauthoryear{{Rodr{\'\i}guez-Ardila}, {Pastoriza},
  {Viegas}, {Sigut}  \& {Pradhan}}{{Rodr{\'\i}guez-Ardila}
  et~al.}{2004}]{ardila04}
{Rodr{\'\i}guez-Ardila} A.,  {Pastoriza} M.~G.,  {Viegas} S.,  {Sigut}
  T.~A.~A.,   {Pradhan} A.~K.,  2004, \mn@doi [\aap]
  {10.1051/0004-6361:20034285}, \href
  {https://ui.adsabs.harvard.edu/abs/2004A&A...425..457R} {425, 457}

\bibitem[\protect\citeauthoryear{{Rodr{\'\i}guez-Ardila}, {Riffel}  \&
  {Pastoriza}}{{Rodr{\'\i}guez-Ardila} et~al.}{2005}]{ardila05}
{Rodr{\'\i}guez-Ardila} A.,  {Riffel} R.,   {Pastoriza} M.~G.,  2005, \mn@doi
  [\mnras] {10.1111/j.1365-2966.2005.09638.x}, \href
  {https://ui.adsabs.harvard.edu/abs/2005MNRAS.364.1041R} {364, 1041}

\bibitem[\protect\citeauthoryear{{Rodr{\'\i}guez-Ardila}, {Prieto}, {Portilla}
  \& {Tejeiro}}{{Rodr{\'\i}guez-Ardila} et~al.}{2011}]{ardila11}
{Rodr{\'\i}guez-Ardila} A.,  {Prieto} M.~A.,  {Portilla} J.~G.,   {Tejeiro}
  J.~M.,  2011, \mn@doi [\apj] {10.1088/0004-637X/743/2/100}, \href
  {https://ui.adsabs.harvard.edu/abs/2011ApJ...743..100R} {743, 100}

\bibitem[\protect\citeauthoryear{{Vanzi}, {Cresci}, {Telles}  \&
  {Melnick}}{{Vanzi} et~al.}{2008}]{vanzi08}
{Vanzi} L.,  {Cresci} G.,  {Telles} E.,   {Melnick} J.,  2008, \mn@doi [\aap]
  {10.1051/0004-6361:20078885}, \href
  {https://ui.adsabs.harvard.edu/abs/2008A&A...486..393V} {486, 393}

\bibitem[\protect\citeauthoryear{{Winge}, {Riffel}  \&
  {Storchi-Bergmann}}{{Winge} et~al.}{2009}]{winge09}
{Winge} C.,  {Riffel} R.~A.,   {Storchi-Bergmann} T.,  2009, \mn@doi [\apjs]
  {10.1088/0067-0049/185/1/186}, \href
  {https://ui.adsabs.harvard.edu/abs/2009ApJS..185..186W} {185, 186}

\bibitem[\protect\citeauthoryear{{den Brok} et~al.,}{{den Brok}
  et~al.}{2015}]{denbrok15}
{den Brok} M.,  et~al., 2015, \mn@doi [\apj] {10.1088/0004-637X/809/1/101},
  \href {https://ui.adsabs.harvard.edu/abs/2015ApJ...809..101D} {809, 101}

\makeatother
\end{thebibliography}





\bsp	
\label{lastpage}

\end{document}